\documentclass[acmsmall]{acmart}
\usepackage{threeparttable}
\usepackage{etoolbox}
\appto\TPTnoteSettings{\footnotesize}
\usepackage{array}
\usepackage{tikz}
\tikzset{
  font=\footnotesize,
  qnode/.style = {rectangle,draw,fill=blue!10},
  dummy/.style    = {circle,draw, fill=red!10}
}
\def\BibTeX{{\rm B\kern-.05em{\sc i\kern-.025em b}\kern-.08emT\kern-.1667em\lower.7ex\hbox{E}\kern-.125emX}}

\makeatletter

\DeclareRobustCommand\citepos
  {\begingroup
   \let\NAT@nmfmt\NAT@posfmt% ...except with a different name format
   \NAT@swafalse\let\NAT@ctype\z@\NAT@partrue
   \@ifstar{\NAT@fulltrue\NAT@citetp}{\NAT@fullfalse\NAT@citetp}}

\let\NAT@orig@nmfmt\NAT@nmfmt
\def\NAT@posfmt#1{\NAT@orig@nmfmt{#1's}}

\makeatother

\setcopyright{acmlicensed}
\acmJournal{PACMHCI}
\acmYear{2020} \acmVolume{4} \acmNumber{CSCW1} \acmArticle{27} \acmMonth{5} \acmPrice{15.00}\acmDOI{10.1145/3392832}

\received{January 2020}
\received[revised]{February 2020}
\received[accepted]{March 2020}

\begin{document}

%
% The "title" command has an optional parameter, allowing the author to define a "short title" to be used in page headers.
\title{Methods for Generating Typologies of Non/use}

%
% The "author" command and its associated commands are used to define the authors and their affiliations.
% Of note is the shared affiliation of the first two authors, and the "authornote" and "authornotemark" commands
% used to denote shared contribution to the research.

\author{Devansh Saxena}
\authornote{The first two authors contributed equally to this research.}
\affiliation{%
  \institution{Marquette University}
  \streetaddress{Cudahy Hall, 1313 W Wisconsin Avenue}
  \city{Milwaukee}
  \state{WI}
  \postcode{53233}
  \country{USA}}
\email{devansh.saxena@marquette.edu}

\author{Patrick Skeba}
\authornotemark[1]
\affiliation{%
  \institution{Lehigh University}
  \streetaddress{322 Building C, 113 Research Drive}
  \city{Bethlehem}
  \state{PA}
  \postcode{18015}
  \country{USA}}
\email{pts217@lehigh.edu}

\author{Shion Guha}
\affiliation{%
  \institution{Marquette University}
  \streetaddress{Cudahy Hall, 1313 W Wisconsin Avenue}
  \city{Milwaukee}
  \state{WI}
  \postcode{53233}
  \country{USA}}
\email{shion.guha@marquette.edu}

\author{Eric P. S. Baumer}
\affiliation{%
  \institution{Lehigh University}
  \streetaddress{235 Building C, 113 Research Drive}
  \city{Bethlehem}
  \state{PA}
  \postcode{18015}
  \country{USA}}
\email{ericpsb@lehigh.edu}

%
% By default, the full list of authors will be used in the page headers. Often, this list is too long, and will overlap
% other information printed in the page headers. This command allows the author to define a more concise list
% of authors' names for this purpose.
\renewcommand{\shortauthors}{Saxena and Skeba, et al.}

%
% The abstract is a short summary of the work to be presented in the article.
\begin{abstract}
 
 Prior studies of technology non-use demonstrate the need for approaches that go beyond a simple binary distinction between users and non-users. This paper proposes a set of two different methods by which researchers can identify types of non/use$^{1}$ relevant to the particular sociotechnical settings they are studying. These methods are demonstrated by applying them to survey data about Facebook non/use. The results demonstrate that the different methods proposed here identify fairly comparable types of non/use. They also illustrate how the two methods make different trade offs between the granularity of the resulting typology and the total sample size. The paper also demonstrates how the different typologies resulting from these methods can be used in predictive modeling, allowing for the two methods to corroborate or disconfirm results from one another. The discussion considers implications and applications of these methods, both for research on technology non/use and for studying social computing more broadly.
 
\end{abstract}



%
% Keywords. The author(s) should pick words that accurately describe the work being
% presented. Separate the keywords with commas.

%
% A "teaser" image appears between the author and affiliation information and the body 
% of the document, and typically spans the page. 
%%  \includegraphics[width=\textwidth]{sampleteaser}
%%  \caption{Seattle Mariners at Spring Training, 2010.}
%%  \Description{Enjoying the baseball game from the third-base seats. Ichiro Suzuki preparing to bat.}

%
% This command processes the author and affiliation and title information and builds
% the first part of the formatted document.
\maketitle

\section{Introduction}

A growing body of arguments and evidence make clear the insufficiency of a binary distinction between ``users'' and ``non-users.'' Prior work describes myriad forms of engagement with and disengagement from technology \citep{Wyatt-2003-NonUsersAlsoMatter,SatchellDourish-2009-userusenonuse,BrubakerAnannyEtAl-2016-Departingglancessociotechnical,Schoenebeck-2014-GivingTwitterLent,BaumerAdamsEtAl-2013-LimitingLeavingRe,BaumerGuhaEtAl-2015-MissingPhotosSuffering,BaumerBrubaker-2017-Postuserism,HarmonMazmanian-2013-StoriesSmartphoneEveryday,Leavitt-2015-ThisThrowawayAccount,SambasivanCutrellEtAl-2010-IntermediatedTechnologyUse,BrubakerCallison-Burch-2016-LegacyContactDesigning}. 

It is less clear, though, exactly what kind of an alternative to a binary distinction researchers should employ. Despite the growing body of research cited above, work in this area has yet to articulate a single, definitive typology of non/use\footnote{Following \citet{BaumerBurrellEtAl-2015-ImportanceImplicationsStudying}, this paper employs the term ``non/use'' as a shorthand for ``use and non-use.''}.

This situation likely arises in part from the fact that the types of relationships (i.e., the subject positions \citep{BaumerBrubaker-2017-Postuserism}) in question are often specific to particular technologies and social contexts. Reddit is unique in its policy condoning and even encouraging throw-away accounts \citep{Leavitt-2015-ThisThrowawayAccount}. Facebook's deactivation mechanism creates a liminal status \citep{BaumerAdamsEtAl-2013-LimitingLeavingRe}. Grindr's location-based nature grants it varying salience depending on a user's geographic location \citep{BrubakerAnannyEtAl-2016-Departingglancessociotechnical}. This small handful of examples helps illustrate the difficulty in developing a single, canonical typology to account adequately for the myriad forms of sociotechnical dis/engagements observed in practice.

This lack of means by which to agree upon a typology for forms of non/use can become an impediment to certain types of research. While qualitative research can illuminate the above described complexity, quantitative research that makes use of inferential statistics such as Generalized Linear Models (GLMs), many times based on analysis of survey data, often requires a finite number of categories to converge successfully \cite{madsen2010introduction}. These categories must also have prescriptive, unambiguous, mutually exclusive criteria by which each participant can be assigned to exactly one category. Such categories almost necessarily gloss over the variety of experiences and relationships that study participants have with technology. Indeed, in trying to analyze a dataset about Facebook non/use, the authors of this paper found themselves in such a situation, caught between the Scylla of over-simplified classification schemes and the Charybdis of rich but uncodifiable collections of practices. We suspect that other researchers studying non/use likely face similar challenges. 

To address these difficulties, this paper contributes a set of two distinct methods for generating typologies of non/use. The first method uses negative binomial curve-fitting to identify the most prevalent types of non/use. The second method generates and assesses different taxonomic trees for assigning a non/use type to each respondent. These methods are demonstrated using survey data related to Facebook non/use. For each method, the paper describes both the process and the rationale for each step therein. The results show how the two methods proposed here generate typologies that are almost directly comparable. These two typologies are then each used to develop separate statistical models for predicting each survey respondent's non/use type. We demonstrate how alignments between the two models' results can serve to increase confidence in those specific results, while divergence can introduce a healthy skepticism. We also compare these results with a previous \textit{a priori} typology from the literature that had not been developed in a data-driven fashion \citep{baumer2019all}. The results show a strong alignment between that \textit{a priori} typology and the data-driven typologies generated here, more so for the taxonomic tree typology than for the curve-fitting typology. The discussion addresses how similar approaches could be adapted not only to examining technology non/use in different contexts, but also to much broader aims of generating typologies of human behavior in a variety of domains.

\section{Related Work}

\subsection{Types of Technology Non/use}

Numerous different approaches have been taken to classifying different types of technology use and non-use. Much of the earliest work around non-use is framed in terms of a digital divide \citep{Wyatt-2003-NonUsersAlsoMatter}. Many studies, especially surveys, continue to treat non-use in a binary fashion \citep{StiegerBurgerEtAl-2013-WhoCommitsVirtual,RyanXenos-2011-WhousesFacebook,Tufekci-2008-GroomingGossipFacebook,Hargittai-2008-WhoseSpaceDifferences,AcquistiGross-2006-ImaginedCommunitiesAwareness,ArchambaultGrudin-2012-LongitudinalStudyFacebook}.

Others, however, have offered approaches that go beyond a simple dichotomy \citep{BaumerAmesEtAl-2015-WhyStudyTechnology,BaumerBurrellEtAl-2015-ImportanceImplicationsStudying}. \citet{LenhartHorrigan-2003-RevisualizingDigitalDivide} argue that the digital divide in terms of internet access should be seen not as a binary but as a spectrum. This spectrum includes, among others, the truly unconnected, who have no possibility for internet access; dropouts and evaders, who could have internet access but choose not to; and home broadband users. While not following the specific types articulated by \citet{LenhartHorrigan-2003-RevisualizingDigitalDivide}, more recent work has similarly considered intensity of use or non-use as a factor \citep{HargittaiHsieh-2010-DabblersOmnivoresTypology,LampeVitakEtAl-2013-UsersNonusersInteractions,WellsLink-2014-FacebookUserResearch}.

\citet{Wyatt-2003-NonUsersAlsoMatter} combines the dimension of volitionality - willingly choosing to forgo technology - with temporality. Doing so differentiates between resisters, who have never adopted a given technology, and rejecters, who previously used the technology but have ceased doing so. \citet{SatchellDourish-2009-userusenonuse} list six different potential types of non-use, from lagging adoption, to disenchantment, to disinterest. This approach emphasizes not only the form but the sociocultural significance of the technology and/or its non-use. Similarly, researchers have started investigating how technology non/use practices vary between different ethnic and socio-economic groups \cite{garg2019you}, with respect to different IoT devices \cite{garg2019analysis}, as well as users' adoption of non-use practices in response to a company's values and/or actions \cite{li2019people}.

\citet{HarmonMazmanian-2013-StoriesSmartphoneEveryday} articulate four different subject positions of use and non-use, ranging from the multi-tasking master to the out-of-touch luddite. They argue that, in practice, many individuals are in a constant state of tension. Continuously negotiating their engagement with and disengagement from technology causes them to fluctuate among these different subject positions. Many of these prior typologies have appealing qualities. For example, practices such as lagging adoption \citep{SatchellDourish-2009-userusenonuse}, resistance \citep{Wyatt-2003-NonUsersAlsoMatter}, or the multi-tasking master \citep{HarmonMazmanian-2013-StoriesSmartphoneEveryday} could conceivably occur with any of a variety of technologies.

However, whether or not each of this practices \textit{actually} occurs is a different question. Experiences such as ``fading away'' \citep{BrubakerAnannyEtAl-2016-Departingglancessociotechnical} have been noted only with certain technologies. Similarly, only some work has analyzed cases of individuals reverting to a technology that they had previously ceased using \citep{BaumerGuhaEtAl-2015-MissingPhotosSuffering,Schoenebeck-2014-GivingTwitterLent}.

Furthermore, many systems have unique technical capabilities for enabling non/use, including limited use. Facebook allows users to deactivate, which hides their profile from other users but retains their data on Facebook's servers for reactivation at any time \citep{Facebook--Howdeactivatemy}. Some users attempt to engage with technology only in ways that they find meaningful \citep{LukoffYuEtAl-2018-WhatMakesSmartphone} or employ one set of technologies to control their use of others \citep{Plaut-2015-Technologiesavoidanceswear,LyngsLukoffEtAl-2019-SelfControlCyberspaceApplying}. Returning to social media after a break is sometimes associated with pruning one's friend list \citep{BaumerGuhaEtAl-2015-MissingPhotosSuffering,Schoenebeck-2014-GivingTwitterLent,BaumerSunEtAl-2018-DepartingReturningSense}. Such unfriending involves different steps and has different social meanings among different social media platforms \citep{Boyd-2006-FriendsFriendstersTop,Gershon-2011-UnFriendMyHeart,KwakChunEtAl-2011-FragileOnlineRelationship,Kivran-SwaineGovindanEtAl-2011-ImpactNetworkStructure}.

This variety makes clear two points. First, documented forms of non/use are both numerous and increasing in number with almost every study. Many of those forms are specific to a given sociotechnical setting or context of non/use. It is also unclear how best to compare results across different studies when such varied approaches are used to identifying non/use types.

Second, researchers lack a rigorous means of determining what counts (or perhaps should count) as a distinct type or form of non/use, either in a specific setting or more generally. Put differently, there is no clearly principled way to determine which non/use practices one should expect to observe in any given sociotechnical system. To echo sentiments from the introduction, asking about every previously-observed form of non/use likely creates distinctions that are either too general to be informative or too fine-grained to be meaningful. Thus, rather than create a single typology that could apply to any situation, this paper instead offers methods by which researchers can develop a typology specific to their research setting.

\subsection{Methods for Typology Development}

Various related bodies of work in HCI, CSCW, and social computing have had differing approaches to thinking about typology development. Some of this variety results from drawing on a varied set of ``parent'' disciplines, such as psychology, sociology, communication, computer science, information science etc. 

In general, typology development can refer to two distinct scientific activities. First, in a broad sense, it may refer to theory building that centers around classifying phenomena in different ways. This is usually (but not entirely) qualitative and draws from existing literature to construct a grounded notion of how an aspect of the world works or is explained via sorting and categorization. For instance, within HCI/CSCW/social computing, typologies drawn from largely qualitative research practices such as focus groups, interviews, ethnographies, and exploratory cluster analysis have been developed in a wide variety of domains such as blockchain applications \cite{elsden2018making}, universal design principles for software applications \cite{fuchs2010hci}, online engagement \cite{white2011visitors}, and social networking site users \cite{brandtzaeg2011typology}. A meta analysis of media-user typology research confirms the above suite of methods for developing typologies as common practices in this broad discipline \cite{brandtzaeg2010towards}. A common critique of these approaches is that they are hard to generalize between and across various domains or context withing HCI/CSCW/social computing \cite{muller2016machine}.

Second, in a more specific sense, typology development may refer to the methodological development of theoretical categories into classification schemes and is usually facilitated via quantitative methods. There exists a wide body of literature within statistics and machine learning for optimizing the number of classes and classification rules. General methods include multi-class support vector machines \cite{weston1998multi},  various ensemble learning practices \cite{dietterich2000ensemble}, and of course, the more recent deep learning methods \cite{chen2014deep}. However, one common critique with the above range of methods is that they engage less with theory and are hard to interpret \cite{doshi2017towards}, especially when developing typologies or classes purely from a mathematical optimization perspective.

This is not to say that efforts have not been made within HCI/CSCW/social computing to reconcile two different methodological approaches to the same problem \cite{muller2016machine, baumer2017comparing}. For instance, very recently, Latent Profile Analysis (LPA) and Latent Class Analysis (LCA) have been proposed to bridge this gap \cite{oberski2016mixture}. However, this approach depends on a thorough knowledge of mixture models which are technically more complex and can be harder to interpret especially when comparing the classes between two typologies \cite{chen2018interpretable}. We examined other scientific domains such as environmental engineering and energy forecasting that have made good use of curve-fitting \cite{gentry2006forecasting} and computational taxonomic methods \cite{kang2009learning} to bridge the qualitative and quantitative method gap described above. We adapt these methods in our study to develop a generalizable and more accessible method (that offers higher transparency and interpretability) for HCI/CSCW/social computing researchers studying complex theoretical constructs.

\section{Case Study: Developing Facebook Non/use Typologies}

The above literature describes varied forms of Facebook (and, more broadly, social technology) non/use. These forms were used to craft the following Yes/No questions as survey items. Following \citet{Wyatt-2003-NonUsersAlsoMatter}, each respondent can be labeled as engaging or not engaging in each of these activities. Parenthetical text, which was not included in the survey, provides additional details relating each item to prior studies.

\begin{itemize}
    \item I \textit{currently} have an active Facebook account.
    \item I have \textit{more than one} Facebook account. (Multiple accounts are sometimes used to manage or limit one's own Facebook use and/or others' access to one's online profile \citep{boyd-2014-ItComplicatedSocial,Marwickboyd-2010-tweethonestlytweet,Marwickboyd-2014-NetworkedprivacyHow}).
    \item At some time, I have \textit{deactivated} my Facebook account. \citep{BaumerAdamsEtAl-2013-LimitingLeavingRe,Schoenebeck-2014-GivingTwitterLent,boyd-2014-ItComplicatedSocial}.
    \item At some time, I have permanently \textit{deleted} my Facebook account.    \citep{BaumerAdamsEtAl-2013-LimitingLeavingRe,Portwood-Stacer-2013-Mediarefusalconspicuous}.
    \item At some time, I have voluntarily \textit{taken a break} from Facebook for a week or more. \citep{BaumerAdamsEtAl-2013-LimitingLeavingRe,RainieSmithEtAl-2013-ComingGoingFacebook,Schoenebeck-2014-GivingTwitterLent,BaumerGuhaEtAl-2015-MissingPhotosSuffering}
    \item At some time, I have used \textit{software} to limit my Facebook usage. \citep{Schoenebeck-2014-GivingTwitterLent,Plaut-2015-Technologiesavoidanceswear}
    \item At some time, I have \textit{deleted the Facebook app} from my phone. \citep{BaumerGuhaEtAl-2015-MissingPhotosSuffering}
    \item I have \textit{never} had a Facebook account. \citep{DugganEllisonEtAl-2015-PewSocialMeida}
    \item It is \textit{my own choice} whether or not I use Facebook. (Prior work has documented where people, against their will, either were prevented from using Facebook or felt forced to have an account \citep{Wyatt-2003-NonUsersAlsoMatter,BaumerAdamsEtAl-2013-LimitingLeavingRe,SatchellDourish-2009-userusenonuse,WycheSchoenebeckEtAl-2013-FacebookLuxuryExploratory,WycheBaumer-2016-ImaginedFacebookexploratory,BaumerSunEtAl-2018-DepartingReturningSense})
\end{itemize}

To be sure, this typology is non-exhaustive. For example, it does not include politically-motivated technology abstention \citep{Portwood-Stacer-2013-Mediarefusalconspicuous}, gradual ``fading away'' of a technology \citep{BrubakerAnannyEtAl-2016-Departingglancessociotechnical}, constant feelings of tension and instability \citep{HarmonMazmanian-2013-StoriesSmartphoneEveryday}, or asking a friend to change one's password \citep{BaumerAdamsEtAl-2013-LimitingLeavingRe}. Instead, it focuses on the most common forms of Facebook non/use documented in prior work that could conceivably be captured with Yes/No questions. The correlation matrix of these questions can be found in the appendix.

A survey design was sought that would encompass as many of these different practices as possible. However, prior literature provided little \textit{a priori} expectation about how these individual practices might be organized into a coherent typology. For example, should deactivating one's account be seen as a special case of taking a break from Facebook? Should it be assumed that all respondents who deleted their account also deleted the Facebook app from their phone? What if a respondent with multiple Facebook accounts deactivated one account, deleted another, and left a third account active?

Instead of making such decisions at the outset, each respondent answered every one of the Yes/No questions above. Not only does this approach obviate the need to make assumptions about relationships among these practices, it also extends \citepos{Wyatt-2003-NonUsersAlsoMatter}. She used two dimensions, temporality (used technology before or did not) and volitionality (non-use is a willful choice or not), resulting in a 2-dimensional grid with a total of $2^2=4$ types of non/use. We extend this idea to generate a N-dimensional grid with a total of $2^T$ types, where $T$ is the number of Yes/No questions included. Researchers studying different data sets or working in other domains could generate their own list of Yes/No questions based on relevant prior literature and expectations, a point considered further below in the Discussion.

The responses to these Yes/No questions were used to generate typologies from the bottom-up. There is little reason to expect that at least one participant will occupy each and every one of the total $2^T$ (in this case, $2^9=512$) types of non/use. Furthermore, some types may never occur; we would expect it impossible that a single participant \textit{both} never had a Facebook account \textit{and} currently has an active Facebook account. Thus, we propose a data-driven approach to identify the most perspicacious typology possible, as described below in Section \ref{sec:AnalysisAndResults}.

Using two different methodological lenses, we derived two typologies based on participants' responses to the seven typological survey questions (both typology generation methods resulted in dropping two of the typological survey questions, as described below). For each of these typologies, we then trained a multinomial logistic regression model to predict non/use type -- one separate model for each typology -- using features extracted from the psychometric scales included in the survey. The goal was to evaluate whether these two different methods would converge towards a single, consistent non/use typology or generate divergent, non-comparable typologies.

\subsection{Data}

Data were collected as part of a larger survey about Facebook non/use. Participants were recruited via Qualtrics, whose recruitment and sampling procedure is outlined on their website (https://www.qualtrics.com/online-sample/). Qualtrics' staff assembled a web panel of participants with a demographic composition resembling that of general internet users with respect to age, education, and ethnicity \citep{PewResearchCenter-2018-DemographicsInternetHome}. Respondents were screened at the beginning of the survey using these criteria. For example, once the survey received 88 respondents age 25-34 (i.e., 17.7\% of our target sample size of 500 respondents), subsequent respondents in the age 25-34 did not pass the age criterion. Respondents who did not pass any of the demographic screening criteria were excluded. Ultimately, the web panel included 516 participants (with 1028 potential respondents screened for not passing demographic criteria), for which we paid \$2,500.

The survey also included a variety of other items that might relate to Facebook non/use, mostly drawn from prior related work. The Facebook Inventory (FBI) \citep{EllisonSteinfieldEtAl-2007-BenefitsFacebookFriends} was used to assess intensity of Facebook usage \citep[cf.][]{HargittaiHsieh-2010-DabblersOmnivoresTypology}. A prior uses and gratifications analysis found five distinct factors in why people use Facebook \citep{LeinerKobilkeEtAl-2018-Functionaldomainssocial}; to combat survey fatigue, the present survey included only the two highest-loading questions for each of these factors. The six-item version of the Bergen Facebook Addiction Scale (BFAS) \citep{AndreassenTorsheimEtAl-2012-DevelopmentFacebookAddiction} was used to assess limited impulse control \citep{Grant-2008-ImpulseControlDisorders,GrantPotenzaEtAl-2010-IntroductionBehavioralAddictions}\footnote{Since there is some debate over the suitability of ``addiction'' to describe potentially problematic social media usage \citep{Griffiths-2013-SocialNetworkingAddiction,GriffithsKussEtAl-2014-SocialNetworkingAddiction,Davis-2012-ProblemInternetAddiction,Portwood-Stacer-2012-HowWeTalk}, BFAS is used here as a measure of a respondent's ability to control their own usage of Facebook, as well as negative consequences that may arise from high levels of usage}. A ten-item scale \citep{GoslingRentfrowEtAl-2003-verybriefmeasure} assessed personality \citep{CostaMcCrae-1992-RevisedNEOPersonality}, which prior work has found to differ in binary comparisons of users and non-users \citep{RyanXenos-2011-WhousesFacebook,StiegerBurgerEtAl-2013-WhoCommitsVirtual}. Since prior work has also found an individual's subjective feelings of control to be an important factor \citep{BaumerSunEtAl-2018-DepartingReturningSense,WisniewskiXuEtAl-2014-UnderstandingUserAdaptation,Schoenebeck-2014-GivingTwitterLent,HarmonMazmanian-2013-StoriesSmartphoneEveryday}, the survey included a validated scale for individual sense of agency \citep{TapalOrenEtAl-2017-SenseAgencyScale}. Since multiple prior studies have linked privacy and non/use \citep{AcquistiGross-2006-ImaginedCommunitiesAwareness,BaumerAdamsEtAl-2013-LimitingLeavingRe,RainieSmithEtAl-2013-ComingGoingFacebook,StiegerBurgerEtAl-2013-WhoCommitsVirtual,Tufekci-2008-GroomingGossipFacebook}, the survey included several questions about disclosure of potentially private information \citep{Jeong2017,Wang2011}. The survey included several self-report questions about respondents' social networks; in the interest of simplicity, these are omitted from the present analysis. Several questions had optional free-text fields for participants to further elaborate a response; for instance, participants who indicated that they had multiple accounts were prompted to explain why they did so. The free-text responses were not used in this quantitative analysis, except to check whether participants were interpreting the questions in the manner we intended. Finally, demographics included gender (male, female, other), household income, zip code, marital status, and political views (7-point Likert from Very Liberal to Very Conservative).

\section{Two Methods of Typology Generation} \label{sec:AnalysisAndResults}

Due to the absence of definitive methodologies for generating typologies of non/use, we opted for a data-driven approach. We considered two methods, based on different statistical techniques, and explain here the tradeoffs that each method entails. In the subsequent section, we present the results of an experiment in which we train a multinomial logistic regression model to predict classes in each of the typologies based on the survey data. This experiment helps to illustrate the value of each typology and encourages us to suggest that future work consider multiple non/use typologies. Such an approach should prove especially useful when, as in the case of Facebook, no definitive classes of non/use exist. We also provide a blueprint (illustrated in Figure \ref{fig:comptyps}) for this typology generation process which will allow researchers working in other domains to replicate our methodology for generating classes.

\subsection{Method 1: Curve-fitting solution}

Broadly speaking, the negative binomial curve fitting approach identifies every unique combination of Yes's and No's across all the typology questions. We dropped two of the nine typological questions because no (or very few) users responded affirmatively to these two questions. \textit{"I have never had a Facebook account"} represents zero users and \textit{"It is my own choice whether or not to use Facebook"} represents less than 2\% of the users.

Each unique combination of Yes's and No's to the remaining seven questions becomes a type of non/use. Our initial analysis of the 7 typological questions revealed 50 unique user-behaviors. These questions were borrowed from prior literature as being significant in identifying types of non/use. However, we sought to avoid any \textit{a priori} assumptions, either about which of these practices do and do not co-occur, or about their relative prevalence. Therefore, we use this data-driven approach to figure out the most prevalent types of non/use. This approach then attempts to identify the optimal number of types to include, balancing the number of respondents who fit into one of those types (more is preferred) against the total number of types (fewer is preferred). For this step, we focused on the top fifteen most common non/use types, as depicted in Table \ref{tab:CurveFittingTypology}. After the top fifteen non/use types, very few participants (less than 6) belonged to the classes that fell in the long tail of the distribution.  

\begin{figure}
  \includegraphics[width=\textwidth]{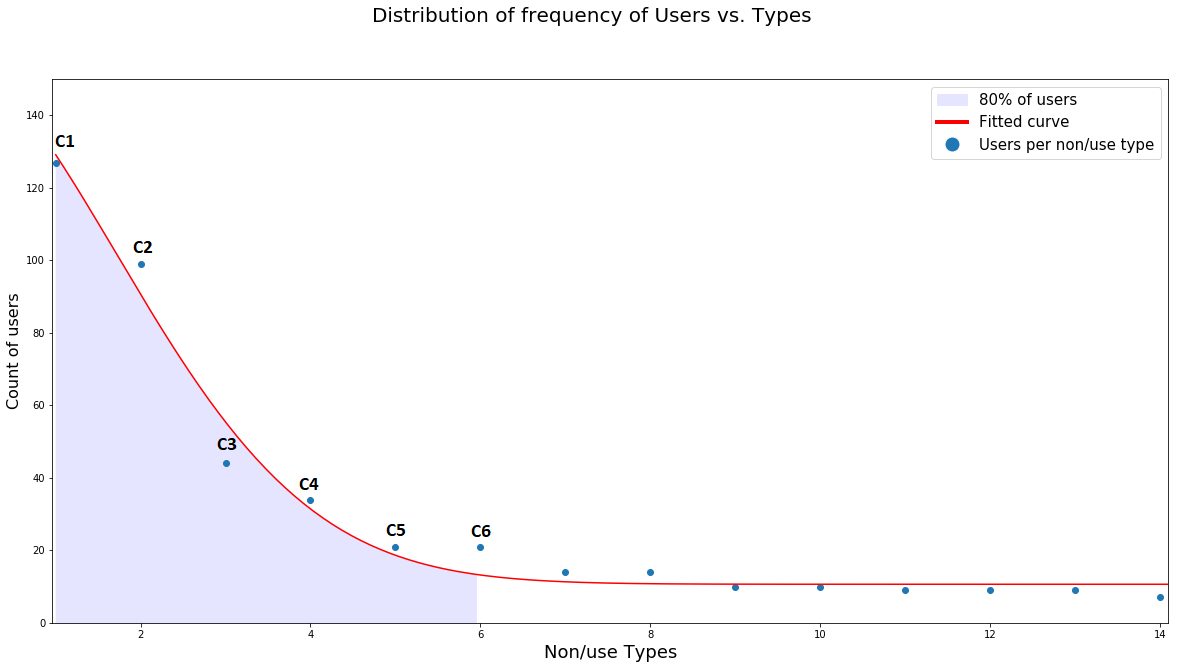}
  \caption{Curve-fitting solution describing non-use behavior}
  \Description{Curve-fitting solution describing non-use behavior}
  \label{fig:CurveFittingPlot}
    \begin{tablenotes}
        \item [1]\textbf{C1}: FB
        \item[2]\textbf{C2}: FB + takenBreak
        \item[3]\textbf{C3}: FB + takenBreak + deactivated + deletedApp 
        \item[4]\textbf{C4}: FB + takenBreak + deletedApp
        \item[5]\textbf{C5}: FB + takenBreak + deactivated
        \item[6]\textbf{C6}: FB + takenBreak + deactivated + deletedApp + deleted
        \item[7]
        \item[8]\textbf{Note}: The "+" sign indicates that the users had engaged in the specified non/use behaviors at some point in the past. The sign does not establish an ordinal or temporal relationship between the non/use behaviors.
    \end{tablenotes}
\end{figure}

Next, we construct a mathematical curve that best fits the relationship between the number of respondents of each non/use type and the frequency of each non/use type in the data set (see Figure \ref{fig:CurveFittingPlot}). A negative binomial distribution is the best fit theoretical distribution for this relationship. Figure \ref{fig:CurveFittingPlot} depicts the fitted mathematical function represented by the following equation where \textit{x} represents each non/use type (ordered by frequency in the data set) and \textit{y} is the number of users corresponding to that non/use type.

\[ y = 10.635 + 150.740e^{-((x+0.604)/3.271)^2}\]

In accordance with the Pareto principle \cite{box1986analysis}, we focus on the non/use behavior of 80\% of the respondents. Our research design uses Pareto's 80/20 rule but it is not a compulsion, and other studies may deviate from this ratio based on the number of classes present on the left side of the equation. Generalized Linear Models are built using maximum likelihood estimation and their performance deteriorates as the number of classes on the left side of the equation increases \cite{madsen2010introduction}. Therefore, it is imperative to limit the number of classes on the left side of the regression model, i.e., the number of distinct types of non/use emerging from the typology. Figure \ref{fig:CurveFittingPlot} shows that the top six non/use types can be used to describe the non/use behavior of 80\% of the respondents.

\begin{table}
    \small
  \begin{tabular}{p{12cm}p{0.5cm}}
    \toprule
       Type & Users\\
    \midrule
        FB & 127\\
        FB + takenBreak & 99\\
        FB + takenBreak + deactivated + deletedApp & 44\\
        FB + takenBreak + deletedApp & 34\\
        FB + takenBreak + deactivated & 21\\
        FB + takenBreak + deactivated + deletedApp + deleted & 21\\
        FB + takenBreak + deactivated + deletedApp + FBmorethan1 & 14\\
        FB + FBmorethan1 & 14\\
        takenBreak + deactivated + deletedApp + deleted & 10\\
        FB + deactivated & 10\\
        FB + takenBreak + FBmorethan1 & 9\\
        FB + takenBreak + deactivated + + deletedApp + usedSoftwareToLimit & 9\\
        FB + takenBreak + deactivated + deletedApp + deleted + usedSoftwareToLimit  & 9\\
        deactivated + takenBreak + deletedApp & 7\\
        FB + deletedApp & 7\\
    \bottomrule
  \end{tabular}
  \vspace{0.1cm}
  \begin{tablenotes}
        \item [1]\textbf{FB}: I currently have a Facebook account 
        \item[2]\textbf{FBmorethan1}: I have more than one Facebook account
        \item[3]\textbf{deactivated}: At some point, I have deactivated my Facebook account 
        \item[4]\textbf{deleted}: At some point, I have permanently deleted my Facebook account
        \item[5]\textbf{takenBreak}: At some point, I have voluntarily taken a break from Facebook for a week or more
        \item[6]\textbf{usedSoftwareToLimit}: At some time, I have used software to limit my Facebook usage
        \item[7]\textbf{deletedApp}: At some point, I have deleted the Facebook app from my phone
        \item[8]
    \end{tablenotes}
\caption{Typology generated by curve-fitting method. The "+" sign indicates that the users had engaged in the specified non/use behaviors at some point in the past. The sign does not establish an ordinal or temporal relationship between the non/use behaviors.}
\label{tab:CurveFittingTypology}
\end{table}

% Begin Patrick's analysis

\subsection{Method 2: Taxonomic-Tree solution}

As a complement to the above curve-fitting solution for determining typologies, one could also use a taxonomic tree to select a subset of typology questions and place all respondents into exactly one category. The basic process is to choose a question on which to divide the population into the `Yes' and `No' groups, and then iteratively subdivide those groups on new questions until either all nine questions are used or one of the subdivisions has too few members, according to some criteria. We chose 40, or approximately 8\% of the 514 total respondents, as the minimum and stopped the iterative splitting if it resulted in groups with less than this number of members.

\textit{Choosing a Taxonomy-based Typology} -- 
The process of selecting a taxonomic tree to define the typologies is nontrivial. Enumeration of all possible trees is known to be NP hard and could become infeasible as the number of questions in consideration grows, necessitating the use of more sophisticated algorithms like the one described in \citep{Ruggieri2017}. In our case, a naive approach was sufficient, but this issue of scalability should be noted. By requiring a minimum of 40 members in each group we restricted the number of possible trees, resulting in a total of 329 candidates.

The second challenge was deciding which of the 329 candidate trees to use in defining the typologies. To our knowledge, the problem of selecting the ``best'' taxonomic trees for use as an outcome variable in a regression model is an unexplored area of research. Thus, we devised evaluation criteria that fit our dataset. First, we excluded all trees which contained a leaf with fewer than 51 members (or 10\% of the total) on the bases that smaller groups would create challenges for the regression models and risk over-fitting. This significantly reduced the number of candidates from 329 to 59.

Next, we applied an upper bound on the size of each group of 172, or one third of the total number of respondents. This number was chosen to ensure that we had at least two splits in the tree, i.e., at least three groups. This step excludes typologies based on a single question, such as simply comparing those respondents who have deleted their account vs. those who have not. Application of this criterion resulted in 20 candidate trees.

Half of the remaining 20 trees included splits based on having multiple accounts. Upon closer inspection of those respondents who answered ``yes'', we saw significant disagreements in how the question was interpreted based on the associated free-text fields. Some respondents considered themselves to have multiple accounts because they deleted theirs and created a new one, others noted that they were forced to make a new account because they could not log in to their prior account. Due to the ambiguity of the question as it was worded, we chose to exclude trees which split the data based on maintenance of multiple accounts.

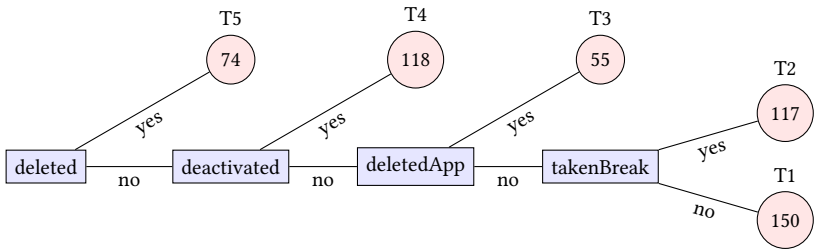
\begin{figure}
    \centering
    \begin{tikzpicture}
  [
    grow                    = right,
    sibling distance        = 4em,
    level distance          = 7em,
    sloped
  ]
  \node [qnode] {deleted}
    child[missing]{node{}}  
    child {node [qnode] {deactivated}
      child[missing]{node{}}    
      child{node [qnode] {deletedApp} 
        child[missing]{node{}}
        child{node [qnode] {takenBreak}
            % child[missing]{node{}}  
            child{node [dummy,label=T1] {150}edge from parent node [below] {no}}
            child{node [dummy,label=T2] {117}edge from parent node [below] {yes}}
        edge from parent node [below] {no}}
        child{node [dummy,label=T3] {55}edge from parent node [below] {yes}}
      edge from parent node [below] {no}}
      child{node [dummy,label=T4] {118} edge from parent node [below] {yes}}
    edge from parent node [below] {no} }
    child { node [dummy,label=T5] {74}
      edge from parent node [below] {yes} }
    ;
    \end{tikzpicture}
    \vspace{0.1cm}
    \caption{Final taxonomic tree used for analysis. Four questions were used to group all 514 subjects into one of 5 typologies, each represented by the red leaf nodes. For example, the second red leaf from the left, T4, corresponds to the 118 participants who responded that they had never permanently deleted their Facebook account but have, at some point, deactivated their account.}
    \label{fig:dectree}
\end{figure}

After removing trees containing the multiple accounts question, we were left with 10 candidates. Finally, we sorted these remaining trees by the variance of the leaves, in order to generate a taxonomy with the most equally sized groups. The final selected typology is depicted in Figure \ref{fig:dectree}.

\begin{figure}[h]
  \includegraphics[width=\textwidth]{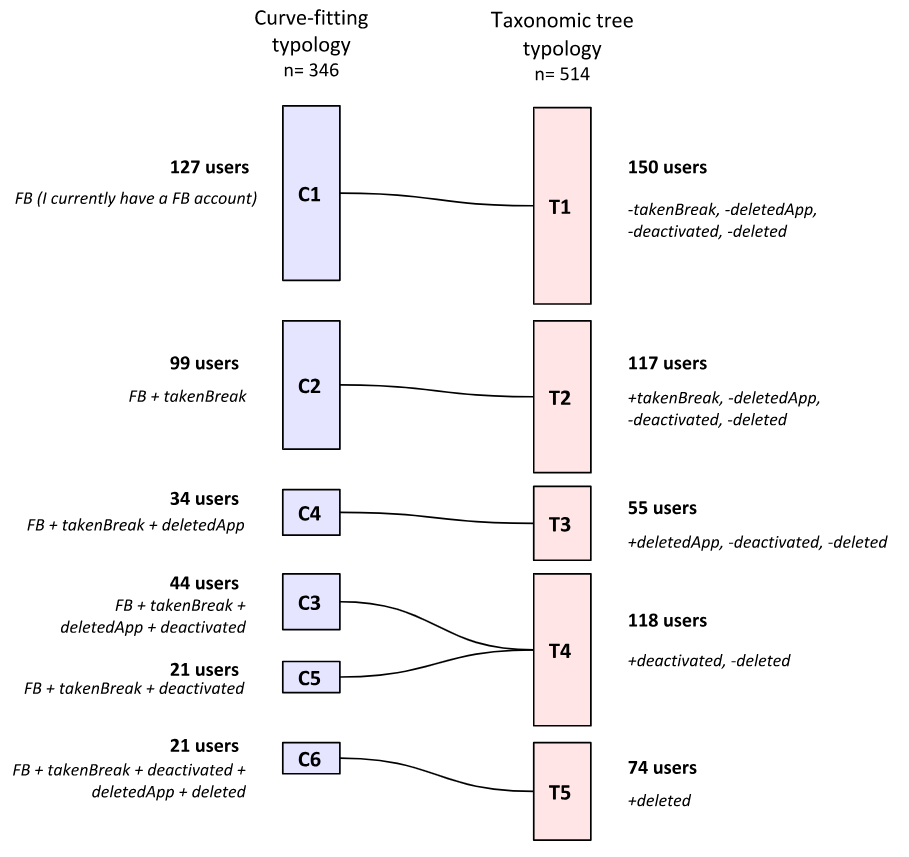}
  \caption{Mapping curve-fitting typology onto the taxonomic-tree typology. The "+" and "-" signs indicate that the users had engaged (or not engaged) in the specified non/use behaviors at some point in the past. The signs do not establish an ordinal or temporal relationship between the non/use behaviors.}
  \Description{Mapping curve-fitting typology onto the taxonomic-tree typology}
\label{fig:comptyps}
\end{figure}

\subsection{Comparing the Two Methods}
The curve-fitting method is an entirely data-driven process that allows us to recognize and model the most prevalent types of non/use while ignoring the more rare types that end up in the long tail of the distribution. Therefore, if researchers were interested in especially studying one of these rare classes, the curve-fitting method would perform poorly because the learning algorithm may fail to identify patterns in the rare classes because of over-representation of the more prevalent classes. For instance, in our survey less than 2\% of the participants responded affirmatively to the question, \textit{"It is my choice whether or not to use Facebook"}. So, researchers who want to study volitionality (non-use is a willful choice or not) would want to use methods (for e.g., taxonomic trees) that allow them to generate typologies that focus on rare classes. Moreover, the curve-fitting method allows us to study the prevalence of non/use behaviors as a general theoretical construct, however, it does not allow us to investigate specific non/use behaviors.

On the other hand, the taxonomic tree method is a data-driven process that allows for qualitative interventions based on the research criteria. In our case, we devised evaluation criteria based on the intensity of non/use behaviors. As mentioned earlier, enumeration of all possible trees in an NP hard problem, therefore, researchers must theoretically design criteria that allows them to generate specific trees that focuses on a specific phenomena. Making these decisions inadvertently means that we will be losing some information since we decided to focus on certain trees that offer the most information about a phenomena. For instance, in our case we had to choose from a subset of the non/use questions that offer the most information with respect to the intensity of non/use behaviors.

%ADD AN EXAMPLE THAT CAN MAKE USE OF BOTH THESE METHODS
These two methods used in conjunction may offer pathways for generating typologies for studying gender identity and sexual orientation. As non-binary genders are becoming increasingly prevalent, it becomes important to be able to study both the prevalence of different genders (curve-fitting) as well as gender variation within non-binary genders (taxonomic trees) without othering or erasing non-binary participants who might associate with types with smaller class sizes \cite{jaroszewski2018genderfluid, spiel2019better, fraser2018evaluating}.

\begin{figure}
  \includegraphics[scale=0.40]{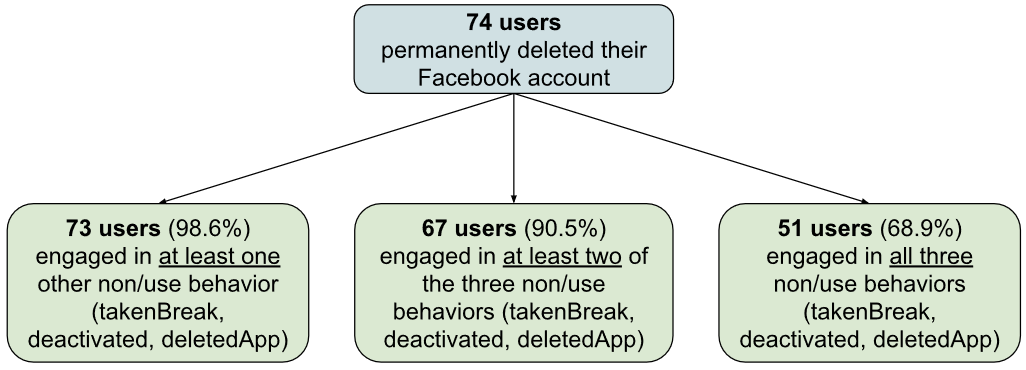}
  \caption{Most participants who permanently deleted their Facebook account (taxonomic type T4) engaged in at least one other non/use behavior (deactivating their account, deleting the app from their mobile device, or taking a break).}
\label{fig:intensity}
\end{figure}

\subsection{Comparing the Two Generation Methods and Typologies}
The most significant consideration when comparing the two methods is the tradeoff between the size of the dataset used for training and the fineness of the groups derived in each typology. The curve-fitting method uses all seven questions to create a class for each unique set of responses, but excludes datapoints in order to discount the long tail and maintain large enough class sizes to make classification feasible. On the other hand, the taxonomic method preserves the number of datapoints, but makes less fine distinctions between classes by only using a subset of the questions. For all classes in the curve-fitting typology, all seven questions are used and the average class size is 58 ($n = 346$), while the taxonomic-tree typology uses a maximum of four questions (T1 and T2) and a minimum of one (T5) with an average class size of 103 ($n = 514$).

In terms of the resultant typologies, a theoretical mapping exists between most of the types. For instance, as depicted in Figure \ref{fig:comptyps}, the curve-fitting type \textbf{\textit{C6}} maps onto the taxonomic-tree type \textbf{\textit{T5}} because participants who engage in the more intense forms of non/use are also more likely to have engaged in the less intense form of non/use. Figure \ref{fig:intensity} illustrates the non/use behaviors of the 74 participants in our study who had permanently deleted their Facebook account (taxonomic type T4). The majority of these participants engaged in other types of non/use, as well. Figure \ref{fig:comptyps} also illustrates the differences in class size and overall number of datapoints, and suggests that the two methods of deriving typologies have in fact converged to similar non/use types. Each taxonomic type has a corresponding curve-fitting type, the latter of which makes up a subset of the former.

\begin{figure}[h]
  \includegraphics[width=\textwidth]{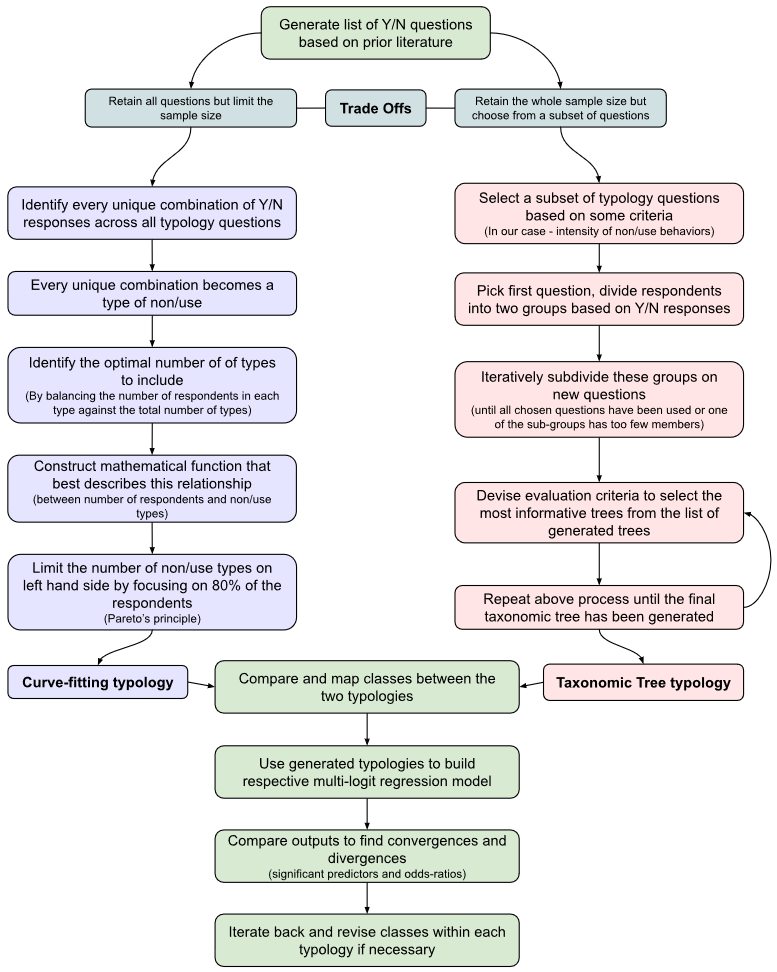}
  \caption{Blueprint for generating typologies for studying and comparing theoretical constructs}
\label{fig:comptyps}
\end{figure}

% Start a new section here
\section{Typology Comparison on Survey Data Classification Task}

Using the typologies described above, we conducted two nearly-identical analyses. Similar to \citet{Baumer-2018-SocioeconomicInequalitiesNon}, both analyses used multinomial logistic regression, where the predictors were attributes of each respondent and the outcome variable was the respondent's non/use type. The only difference was whether each respondent's non/use type (i.e., the outcome variable) was assigned based on the curve-fitting typology or on the taxonomic-tree typology.

This analysis involved a series of steps. First, predictors were extracted from the survey data by performing factor analysis to reduce the dimension of the psychometric scales and including demographic information. Second, model selection was then used to limit predictors to only those which significantly improve our information about the outcome variable. Finally, to compare the two typology generation methods, we identified points of alignment and of divergence, both in terms of which predictors are maintained after model selection, and in terms of the direction and magnitude of those predictors' effects. The remainder of this section describes each of these steps in detail.

\subsection{Feature Extraction from Psychometric Scales}
With either the curve-fitting typology or the taxonomic-tree typology serving as the left side of the regression model (i.e., the outcome variable), we used the five psychometric scales included in the survey to extract features for the right side. For four of the scales, exploratory factor analysis (EFA) was used to identify underlying structures and restrict the number of predictors in the final model. We did not perform factor analysis on the fifth scale, the 10-item Big 5 Personality Scale, choosing instead to use its intended scoring procedure to create a predictor for each of the five personality dimensions. This decision was made after observing that the results from our own factor analysis on the Big 5 items did not match the standard interpretation, which has been validated on large datasets and across different demographics and cultures~\citep{John2008,John1991,Benet-Martinez1998}.

All exploratory factor analyses were performed on the entire 514 member dataset in order to extract the same features with which to train a classification model for both the curve-fitting and taxonomic typologies. Predictors were constructed for each psychometric scale using an average of all questions loading on a given factor, weighted according to the factor loadings. Cronbach's alphas are provided to show internal consistency of factors, and the full factor structures can be found in the appendix.

\subsubsection{Bergens Facebook Addiction Scale (BFAS)}
Andreassen et al in \citep{AndreassenTorsheimEtAl-2012-DevelopmentFacebookAddiction} found six underlying factors using the original 18-question version of the scale which correspond to six dimensions of addictive behavior: salience (the behavior dominates one's thinking and behavior), tolerance (more of a behavior/activity is needed to achieve desired effects), mood modification (activity affects/improves how one is feeling), withdrawal (unpleasant feelings when activity is halted or reduced), conflict (activity causes interpersonal conflict or interferes with other activities), and relapse (repeated reversions to past behaviors after limiting or abstaining)~\citep{AndreassenTorsheimEtAl-2012-DevelopmentFacebookAddiction,Griffiths2005}. Our analysis suggested three underlying factors for this dataset: a combination of Salience, Tolerance, and Mood ($\alpha=.83$) , a combination of Withdrawal and Conflict ($\alpha=.84$), and Relapse. 73\% of the cumulative variance is explained by these three dominant factors.

\subsubsection{Facebook Intensity Scale (FBI)}
Factor analysis was performed on the six FBI questions related to how a person feels about their Facebook use. The remaining two questions on number of Facebook friends and time spent on Facebook were not included in the factor analysis and were treated as individual predictors in the regression models. We found that three factors, which we call \textit{Connectedness} ($\alpha=.86$), \textit{Daily Routine} ($\alpha=.93$) and \textit{Pride}, explains 81\% of the cumulative variance in the full dataset. 

\subsubsection{Sense of Agency (SoA)} 
EFA on the Sense of Agency survey questions suggested two factors, \textit{Sense of Positive Agency} (SoPA) ($\alpha=.79)$ and \textit{Sense of Negative Agency} (SoNA) ($alpha=.84$), which align with the factors identified in \citep{TapalOrenEtAl-2017-SenseAgencyScale}. 43\% of the cumulative variance is explained by these two dominant factors and did not significantly improve with more factors. 

\subsubsection{Leiner et al.'s Uses and Gratifications}
Table \ref{tab:fa:leiner} depicts the factor analysis conducted using questions from \citet{LeinerKobilkeEtAl-2018-Functionaldomainssocial}. We see \textit{Social integration} ($\alpha=.91)$, \textit{Affective gratification} ($\alpha=.91)$, \textit{Personal integration} ($\alpha=.84)$ and \textit{Escape} ($\alpha=.81)$ emerge as the dominant factors. These factors were identified based on the major items referenced in \citet{LeinerKobilkeEtAl-2018-Functionaldomainssocial}. 68\% of the cumulative variance is explained by these four dominant factors.

\subsection{Predicting Curve-Fitting and Taxonomic-Tree Typologies with Multinomial Logistic Regression}

\begin{table}
\Small
\begin{tabular}{|>{\raggedright}p{1.6cm}|>{\raggedright}p{1.8cm}>{\raggedright}p{1.8cm}>{\raggedright}p{2.6cm}>{\raggedright}p{1.5cm}>{\raggedright\arraybackslash}p{2cm}|}
    \toprule
       Y = & BFAS \hspace{0.5cm} + & Leiner \hspace{0.5cm} + & Big5 \hspace{0.5cm} + & Agency \hspace{0.2cm} + &  FBI\\
    \midrule
        \textit{Curve-fitting Typology (C1-C6)} & SalTolMood & Social integration & Extraversion & SoPA & Connectedness\\[15pt]
        & WithdrConfl & Personal integration & Conscientiousness & SoNA & Daily Routine\\[15pt]
        or & Relapse & Escape & Openness &  & Pride\\[15pt]
        \textit{Taxonomic-tree Typology (T1-T5)}  & & Affective gratification & Neuroticism & & FBI-Time\\[15pt]
         & & & Agreeableness & & FBI-Friend\\
         & & & & &\\
    \bottomrule
  \end{tabular}
\caption{Psychometric scale predictors. Each column is based on the factor structure extracted in Section 4.2.}
\label{tab:allpreds}
\end{table}

Finally, we used the predictors derived from EFA on the psychometric scales and demographic items to train two multinomial logistic regression models, one that predicts non/use type according to both the curve-fitting typology (C1-C6), and one that uses the taxonomic-tree typology (T1-T5). In order to avoid overfitting and identify only predictors which significantly increase our knowledge about the outcome variables, we performed model selection to reduce the number of predictors.

\subsubsection{Model Selection: Backward Step-wise Regression}
We conducted step-wise model selections based on AIC for both the curve-fitting and taxonomic-tree typologies. We used a backward elimination process where we start with all predictors depicted in Table \ref{tab:allpreds} and iteratively eliminated predictors whose removal caused the most statistically insignificant deterioration of the model fit. This process helped us identify the most salient predictors for the multi-logit regression models that will be used to classify users into non/use behaviors. Table \ref{tab:reducedpreds} depicts our final predictors for the curve-fitting and taxonomic-tree models along with the odds-ratios for all predictors corresponding to each typology.

\subsubsection{Model Alignments and Divergences} As expected, the reduced models align on some predictors while diverging on others. In terms of alignment, \textit{Relapse} and \textit{Age} are maintained in both models and exhibit similar effects, both in terms of magnitude and in terms of direction. For example, in both models, older respondents were less likely to be anything other than a current user. In terms of divergence, \textit{Daily Routine} and \textit{Affective Gratification} are dominant predictors in both models, but their odds-ratios suggest that they have opposite impacts. With the curve-fitting typology, increases in either of these variables predicted \textit{increased} probability of a respondent being something other than a current user. In contrast, with the taxonomic-tree typology, increases in either of these variables predicted \textit{decreased} probability of a respondent being something other than a current user. Furthermore, five predictors are retained only with the curve-fitting model, and four predictors are retained only for the taxonomic-tree model. The following discussion interprets these alignments and divergences, situating them in related work on social media use and non-use.

\begin{table}[]
\Small
\begin{tabular}{l|l|lllll|llll}
\toprule
Scale & Predictor & \multicolumn{5}{l|}{Odds-Ratios (Curve-fitting)} & \multicolumn{4}{l}{Odds-Ratios (Taxonomic tree)} \\
 &  & C2 & C3 & C4 & C5 & C6 & T2 & T3 & T4 & T5 \\
 \midrule
Agency & SoPa & --- & --- & --- & --- & --- & 0.93** & 1.04** & 1.08** & 1.05** \\
BFAS & Relapse & 1.38* & 1.63* & 1.59* & 1.23* & 1.81* & 1.54*** & 1.93*** & 1.73*** & 1.27*** \\
 & SalTolMood & --- & --- & --- & --- & --- & 0.97 & 0.82 & 0.91 & 1.11 \\
 & WithdrConfl & --- & --- & --- & --- & --- & 0.81 & 0.94 & 1.07 & 1.12 \\
%Big5 & Neuroticism & --- & --- & --- & --- & --- & 0.97* & 0.87* & 0.87* & 0.86* \\
FBI & Connectedness & --- & --- & --- & --- & --- & 1.21 & 1.04 & 0.93 & 0.90 \\
 & Daily Routine & 1.29 & 1.30 & 1.36 & 1.39 & 1.46 & 0.68*** & 0.80*** & 0.78*** & 0.66*** \\
 & Pride & 1.24** & 2.54** & 1.48** & 1.55** & 1.23** & --- & --- & --- &  \\
 & FBI-Friend & 1.05** & 1.72** & 1.28** & 1.07** & 1.40** & --- & --- & --- & --- \\
Leiner & Affective Gratif. & 1.35*** & 1.75*** & 1.65*** & 1.34*** & 1.74*** & 0.93* & 0.88* & 0.88* & 0.84* \\
 & Escape & 0.89* & 0.67* & 0.82* & 0.90* & 0.73* & --- & --- & --- & ---\\
 Demog & Age & 0.98*** & 0.95*** & 0.94*** & 0.97*** & 0.96*** & 0.98*** & 0.94*** & 0.93*** & 0.94*** \\
 \bottomrule
\end{tabular}
\caption{Odds ratios for reduced multi-logit models with both curve-fitting and taxonomy typologies as the classification target. Backwards step-wise regression reduced the curve-fitting model to seven predictors, and the taxonomy model to eight. Reference classes are C1 and T1, which both represent current Facebook users who report no non-use behaviors. Asterisks represent significance level of likelihood-ratio tests on the corresponding predictor and typology ($.05, .01, .001$)}
\label{tab:reducedpreds}
\end{table}

\section{Discussion and Conclusion}
As this is a methodological paper, the discussion focuses primarily on how the methods proposed here can generate useful insights. The contribution is not necessarily the specific insights \textit{per se} into Facebook non/use that arise from the classification models. Rather, the contribution is a demonstration of how these two typology generation methods can be leveraged in combination to produce informative findings.

 Thus, this discussion first considers implications of these typology methods, and how investigating points of alignment and of divergence in the models can lead to valuable observations. Second, we note important limitations that other researchers should bear in mind when using the methods proposed here. Third, the discussion demonstrates that these methods can in fact produce valuable insights by describing the implications of this work for research on technology non/use. Finally, it considers broader implications for the use of quantitative methods to study technology and social behavior.

\subsection{Demonstrating Alignment and Divergence Across Models}

Rather than viewing the curve-fitting and taxonomic tree approaches above as competitors between which one must choose, we instead present a holistic methodological approach that compares results across both typology generation approaches (see Figure \ref{fig:comptyps}). This section draws on Table~\ref{tab:reducedpreds}, which depicts the odds-ratios for the reduced multi-logit models. Recall that the reference categories, C1 and T1, correspond to current Facebook users who have not engaged in any of the non/use practices included in the survey. Therefore, odds-ratios represent the relative probability of a respondent belonging to one of the non/use categories (deletion, deactivation, deleting the app, etc.), compared to this baseline. Examining these odds ratios identifies points of alignment between the two models, reinforcing our confidence in the results. It also draws attention to points of divergence, which can introduce healthy skepticism about single results.

For example, \textit{Relapse} (returning to old habits after a period of abstinence or reduced Facebook use) shows clear alignment across the two models. The corresponding odds-ratios are greater than one and of a similar magnitude across all classes in both typologies. That is, higher reported \textit{Relapse} predicts a higher probability of a respondent engaging in some form of non-use behavior. The similarities in effects and significance (curve-fitting: $p < .05$, taxonomic-tree: $p < .001$) increase our confidence in the predictive power of this feature. These findings also align with prior work on taking breaks from social media and returning before one had intended \citep{BaumerGuhaEtAl-2015-MissingPhotosSuffering,Schoenebeck-2014-GivingTwitterLent,RainieSmithEtAl-2013-ComingGoingFacebook}. Furthermore, \textit{Age} was maintained for both taxonomies with very high significance ($p < .001$); odds-ratios suggest that age is negatively correlated with non-use behaviors (non-adopters were not present in this particular dataset), echoing \citet{Baumer-2018-SocioeconomicInequalitiesNon}. These external resonances lend further credence to our claim that alignment between the two models should increase researchers' confidence in the results.

On the other hand, there are at least two ways that the results from the two typology generation methods suggested here may diverge from one another. First, model selection may cause certain predictors to be maintained in a model with one typology, but not with the other. \textit{Sense of Positive Agency}, for example, is maintained with high significance ($p < .01$) only in the taxonomic-tree model, while \textit{Pride}, \textit{FBI-Friend}, and \textit{Escape} are only preserved for the curve-fitting taxonomy. Inclusion of the two BFAS factors besides \textit{Relapse} slightly reduces AIC in the taxonomic-tree model, but low significance scores cast doubt on their importance. Second, the same predictor across the two different models may have odds-ratios in opposite directions. \textit{Daily Routine} and \textit{Affective Gratification} (described above) both exhibit this phenomenon. That said, differences in significance scores may lend more credence to the effects suggested by one model over those suggested by the other. For example, although \textit{Daily Routine} is maintained in both models, its effect is statistically significant only for the model based on the taxonomic tree typology.

Divergence can be interpreted in a number of different ways. First, these predictors may not actually be significant. That is, their inclusion in the final model may occur only due to statistical chance. Second, the divergences might be resolved with a larger dataset, reducing our uncertainty about the relevancy of certain predictors. Finally, these divergences may suggest that the typologies themselves are not as similar as they initially appear. Consider, for example, the relationship between T4 in the taxonomic tree, and C3 and C5 in the curve-fitting typology (Figure~\ref{fig:comptyps}). Both curve fitting classes (C3 and C5) are strict subsets of the taxonomic tree class (T4). However, C3 distinguishes itself from C5 by requiring that the respondent also deleted the Facebook app from their phone. T4 makes no such distinction. The results show that, when including this distinction within the typology, odds ratios for certain predictors (e.g., \textit{Affective Gratification}, \textit{Escape}) have opposite directions. Thus, the results question whether a respondent deleting the Facebook app from their phone should be used as a decision criterion in classifying their non/use type. More significantly, perhaps the two sets of classes arising from the two different typologies are not quite as similar as depicted in (Figure~\ref{fig:comptyps}). Other researchers using the methods proposed here can similarly identify potentially problematic classes in the typologies generated from their data.

Overall, identifying points of divergence in the models draws our attention to important issues that may have been missed if only one typology were used. Furthermore, doing so raises many questions that can be used to develop more robust typologies of non/use in future work.

At a higher level, such comparisons underscore the strength of the methods proposed here. To reiterate, researchers should not adopt only one of either the curve-fitting or the taxonomic approaches. Rather, they should generate typologies using both these approaches, which enables direct comparisons of the typologies themselves. Furthermore, conducting subsequent analyses, such as the predictive modeling presented here, can provide stronger evidence in support of conclusions drawn from the analysis. To wit, p-values derived from singular studies are noisy, and we hypothesize that adopting a dual methodological approach would yield results that provide more stable outcomes with higher effect sizes \cite{kay2016special}. Issues concerning statistical transparency have received increasing attention in HCI and sociotechnical research \cite{cairns2007hci, bernstein2011trouble}. This work speaks to these issues and offers an example of methodologies that provide greater confidence in our results. We expand on this in a subsequent subsection.

\subsection{Comparison against a third set of non/use classes}
We also wanted to see how our typologies compared to other non/use typologies published in recent literature. \citet{baumer2019all} presented a non/use typology that compares three different forms of non/use with respect to current Facebook users. The three classes discussed in the paper are: (1) current active users; (2) people who deactivated their account; and (3) people who have considered deactivating their account but have not actually done so. We term these B1, B2, and B3, respectively.

A direct comparison with the \citet{baumer2019all} typology was not possible, since our study did not incorporate a non/use question that focused on ``considered deactivation.'' Thus, we compare type B3 with the ``taking a break'' question in our data, which is used in types C2 and T2. Again, this does not provide a direct comparison, since actually taking a break involves action, while considering deactivation does not. However, all of B3, C2, and T2 represent the least intense forms of non/use within their respective typologies. Furthermore, in each case, the user continues to maintain an active channel to their Facebook account. We followed the methodology specified by \citet{baumer2019all} to assign non/use types (B1, B2, or B3) to our participants. We then trained a multinomial logistic regression model followed by a backward step-wise regression model. Table \ref{tab:BaumerTypes} depicts the significant predictors and their odds-ratios determined by the reduced multi-logit model. 

Similar to the curve-fitting and taxonomic-tree typologies, \textit{Relapse} and \textit{Age} are significant predictors for Baumer et al.'s typology, with odds-ratios exhibiting similarity in both magnitude and direction. \textit{Daily Routine} and \textit{Affective Gratification} are also significant predictors, although the magnitude and direction of their odds-ratios aligns with the taxonomic-tree typology and not with the curve-fitting typology. These statistical differences point to the subtle differences that exist between the curve-fitting and taxonomic-tree typologies as stated previously. More importantly, such comparisons allow us (and other researchers) to iterate back and revise the types within each typology to ensure a more robust alignment. 

\begin{table}[]
    \Small
    \centering
    \begin{tabular}{l|l|ll}
         \toprule
         Scale & Predictor &  \multicolumn{2}{l}{Odds-Ratios (Baumer et al. non/use types)} \\
         & & B2 & B3 \\
         \midrule
         BFAS & Relapse & 1.37*** & 1.52*** \\
         FBI  & Daily Routine & 0.73** & 0.78**\\
         Leiner & Affective Gratif. & 0.89* & 0.82* \\
         Demog & Age & 0.97*** & 0.93*** \\
         \bottomrule
    \end{tabular}
    \begin{tablenotes}
        \item[1]\textbf{\hspace{2.4cm} B1}: I currently have a Facebook account (reference category)
        \item[2]\textbf{\hspace{2.4cm} B2}: At some point, I have considered deactivating my Facebook account
        \item[3]\textbf{\hspace{2.4cm} B3}: At some point, I have deactivated my Facebook account
    \end{tablenotes}
    \vspace{0.2cm}
    \caption{Odds ratios for reduced multi-logit model with Baumer et al.'s non/use types \cite{baumer2019all} as the classification target. Backwards step-wise regression reduced the model to four predictors. Asterisks represent significance level of likelihood-ratio tests on the corresponding predictor ($.05, .01, .001$)}
    \label{tab:BaumerTypes}
\end{table}

\subsection{Limitations}
First and foremost, this paper has only examined a single data set, and that data set only pertained to non/use of Facebook. The methods presented here may be more (or perhaps less) difficult with other social media. Moreover, this study only focuses on a subset of non/use types that could conceivably be captured with Yes/No questions (i.e., discrete non/use actions). Furthermore, by focusing on a single social media platform, we were unable to investigate the broader ecology of non/use where users have a choice of several social media platforms \cite{grandhi2019stay}. That said, the relative consistency of the results generated here suggest that the process is fairly robust for generating and investigating non/use typologies. We hope that future work will both apply and adapt the methods offered here.

Second, this method focuses purely on survey data and only uses binary Yes/No questions. As described above, this choice was informed by foundational work on different forms of non-use \citep{Wyatt-2003-NonUsersAlsoMatter}. Future work could expand these methods to include questions with ordinal or continuous responses. Similarly, the methods could be expanded to include data from usage logs \citep{DumaisJeffriesEtAl-2014-UnderstandingUserBehavior}\footnote{In the case of Facebook or other private social media companies, only a limited number of researchers will have access to the details of such log data.}. Again, this paper lays the groundwork upon which future researchers can pursue such directions.

Third, generating novel typologies may significantly limit the ability to compare results across studies. If every data set is used to generate a one-off typology, it becomes difficult to compare and contrast results using such diverse typologies. As an alternative, it might help to consider the methods proposed here as analogous to the construction of validated scales \citep{Furr-2011-ScaleConstructionPsychometrics,Edwards-1983-TechniquesAttitudeScale}. Rather than generate an entirely new typology for every data set, researchers could instead follow an iterative process of development and validation across multiple studies. Such processes occur in numerous domains for the development of standardized, validated scales, from personality \citep{McCraeCosta-1987-Validationfivefactormodel,McCraeCosta-1997-Personalitytraitstructure,GoslingRentfrowEtAl-2003-verybriefmeasure} to privacy \citep{MalhotraKimEtAl-2004-InternetUsersInformation,SmithMilbergEtAl-1996-InformationPrivacyMeasuring,BuchananPaineEtAl-2007-Developmentmeasuresonline}. Similarly, different researchers across multiple studies should test, question, and revise typologies generated in the manner described in this paper.

Finally, there are two general limitations of this type of survey study that deserve mention. First, we can only capture participants' opinions at one instant in time and rely on their recall of past behaviors and characteristics of their Facebook use. This limits our ability to understand temporality, or changes in non/use over time. Other studies have attempted to address such aspects using, e.g., repeated sampling of the same population \citep{BaumerXuEtAl-2017-WhenSubjectsInterpret} or more longitudinal diary studies \citep{BaumerSunEtAl-2018-DepartingReturningSense}. Future work might benefit from combining those techniques with the typology generation methods offered here to examine how typologies themselves change and evolve over time. Second, survey non-respondents may potentially bias our results away from people who are simply less likely to engage in online interaction. This issue similarly occurs in other work targeting individuals who are less directly engaged with technology, such as blog readers \citep{BaumerSueyoshiEtAl-2008-ExploringRoleReader} or lurkers \citep{NonneckePreece-2001-Whylurkerslurk,PreeceNonneckeEtAl-2004-topfivereasons}. We attempted to mitigate this issue by recruiting internet users in a demographically representative sample, rather than Facebook users \textit{per se}. However, this is likely a fundamental issue with survey research, and other methods would be needed to address it more fully.

\subsection{Implications for Technology Non/use Research}

This analysis offers multiple specific implications for research on technology non/use. First, at least in the case of Facebook, it might not make sense to consider volitionality as a factor for identifying types of non/use. As demonstrated in the analysis above, responses to the question (about whether respondents felt they had a choice in whether or not to use Facebook) were ultimately not informative. Initially, this result seems to run contrary to prior work arguing for the analytic importance of an individual's control (or sense of control) over their own technology use \citep{Wyatt-2003-NonUsersAlsoMatter,GuhaBaumerEtAl-2018-RegretsveHad,BaumerGuhaEtAl-2015-MissingPhotosSuffering,WisniewskiXuEtAl-2014-UnderstandingUserAdaptation,SatchellDourish-2009-userusenonuse,WycheSchoenebeckEtAl-2013-FacebookLuxuryExploratory}. However, these results do not suggest that volitionality, or sense of agency, is unimportant. Rather, they suggest that this factor might be less informative for \textit{differentiating} among various types of Facebook non-use and more informative as a \textit{predictor} for an individual respondent's non/use type. Future work will need to examine the utility of this factor for generating typologies in other sociotechnical contexts.

Second, the taxonomic-tree approach suggest some degree of hierarchicality among Facebook non/use types. The most suitable taxonomic-tree prioritizes questions about specific practices in an order that seems to go from most to least intense forms of non-use: from account deletion (permanent and irrevocable), to account deactivation (temporary), to deleting the mobile app (limiting only one access channel), to taking a break (using no technical mechanism). 

Such a hierarchy makes intuitive sense and could have been postulated in advance, based on theory and prior work. However, numerous varied hierarchies could be similarly postulated, with little means of choosing from among them. For example, the types of non/use could have been organized in the opposite direction, i.e., from least to most intense forms of non-use. Doing so, however, would have resulted in severe class imbalances, which would have made predictive modeling difficult. Thus, the choice to avoid an \textit{a priori} hierarchy among different non/use practices allowed our data-driven approach to reveal a latent hierarchy with a meaningful interpretation. Future work should consider whether similarly interpretable typologies emerge from other data sets.

Finally, classification systems embody numerous assumptions, value commitments, power dynamics, etc. \citep{BowkerStar-1999-SortingThingsOut}. The typologies that result from the methods presented in this paper similarly embody commitments and assumptions. Most immediately, the researchers constructing the original Yes/No questions on which the typologies are based have significant influence in what counts as an important factor. More subtly, both the curve-fitting approach and the taxonomic-tree approach favor typologies with fewer types and more respondents in each type. Doing so necessarily marginalizes minority types. For instance, the < 2\% of respondents who indicated that they do \textit{not} have a choice about their Facebook use or non-use may be highly analytically interesting, for reasons described above. However, their severe sparsity in the data set makes them invisible in the typologies generated above. Such concerns have increasingly important implications when methods such as these are applied to the a broader range of typology generation tasks.

\subsection{Implications for Quantitative Methodological Development}

Our study has a number of implications for methodological development in HCI/CSCW/social computing. First, we adapted and implemented two different methods (curve-fitting and taxonomic-trees) for quantitative development of typologies that have been used with some success in other domains such as environmental engineering \cite{gentry2006forecasting} and energy forecasting \cite{kang2009learning}. Our results broadly suggest that these methods can also be used in the generation of typologies around social data to operationalize complex, theoretical constructs. Prior work \cite{brandtzaeg2011typology, elsden2018making, fuchs2010hci, white2011visitors} suggests that the overwhelming majority of work in typology development has been done through theory development and qualitative methods which don't tend to work very well in building statistical models. Specifically, we allude to GLMs (i.e. linear regression, logistic regressions, multi-level models etc.) that are often built to classify typologies with specific sets of predictors. As the number of predicted classes (types) \cite{madsen2010introduction} increase, it becomes harder for the maximum likelihood estimator (MLE) algorithm upon which most GLMS are built to calculate an optimum solution. Thus, optimizing the number of predicted classes is an important consideration that purely qualitative methods may not be able to solve.
Our approach aims to provide a balance between these two tradeoffs. In making this claim, we want to highlight the importance of qualitative methods in typology and theory development but also want to acknowledge that these results are often at odds with the current state of statistical practice \cite{cairns2007hci, bernstein2011trouble, kay2016special, olson2014ways} in HCI/CSCW/social computing.    

Second, our work can be placed in context with the existing state of statistical practices in HCI/CSCW/social computing. Broadly, quantitative researchers in the field either come from a social science background (with requisite expertise in inferential statistics and generalized linear models) or from a computer science background (with general expertise in machine learning) \cite{cairns2007hci, kay2016special}. Unfortunately, neither academic background  lends itself well to rigorous statistical inference, which has rightfully caused researchers to raise concerns in how quantitative methods are developed and adapted in the community \cite{cairns2007hci, bernstein2011trouble, kay2016special}. For instance, it is relatively well established that dependence on noisy p-values and Null Hypothesis Significance Testing (NHST) leads to research that is less statistically rigorous and replicable across different populations and contexts \cite{kay2016special}. This is especially true when operationalizing complex theoretical constructs (such as in our case, non/use) for statistical modeling and inference. Our study suggests that there is space for deeper exploration of methodological development of statistically rigorous, yet accessible methods to the general HCI/CSCW/social computing researcher. For instance, one strong finding from our results suggests that complex theoretical constructs may be conceptualized through multiple predictive modeling approaches to yield greater confidence in patterns, trends and statistical inferences that actually exist in the data (i.e. true positives) and less confidence in those that do not (i.e. false positives or false negatives arising from randomness or noisiness of the data). This type of holistic modeling can be used to ascertain external validity of complex theoretical constructs across samples, populations and contexts beyond the current norm in the field.

Finally, in recent years, HCI/CSCW/social computing venues have been wrestling with the question of what constitutes methodological contribution to the field and whether there exists a set of methods unique to, or particularly amenable for, common questions that arise in HCI/CSCW/social computing \cite{baumer2017comparing, kay2016special, muller2016machine, olson2014ways}. Our work contributes to the conversation in this space by expanding the domain of methodological contribution to typology development and by questioning the current state of statistical reasoning in examining complex, theoretical constructs. With the ever-changing nature of social media and non/use practices, statistical methods such as those presented in this paper may provide consistent, adaptable means to categorize and study these diverse phenomena.

\section{Acknowledgments}
This work is supported in part by the US National Science Foundation (Grant No. IIS-1421498 and Grant No. CNS-1814533) and the Facebook Computational Social Science Methodology Research Award. Thanks to our survey participants for sharing their data and their experiences.

%
% The next two lines define the bibliography style to be used, and the bibliography file.
\bibliographystyle{ACM-Reference-Format}
\bibliography{epsbLibrary,sample-base}
\newpage
\appendix

\section{Correlation matrix for non/use questions}

\begin{table}[h]
    \centering
    \Small
    \begin{tabular}{>{\raggedright}p{2cm}ccccccc}
    \toprule
       Question & Typology-1 & Typology-2 & Typology-3 & Typology-4 & Typology-5 & Typology-6 & Typology-7\\
                & & & & & & & \\    
    \midrule
        Typology-1 & 1.00 & 0.09 & -0.19 & -0.33 & -0.04 & 0.00 & -0.17 \\
        Typology-2 & 0.09 & 1.00 & 0.12 & 0.06 & 0.04 & 0.02 & 0.07 \\
        Typology-3 & -0.19 & 0.12 & 1.00 & 0.45 & 0.37 & 0.16 & 0.55 \\
        Typology-4 & -0.33 & 0.06 & 0.45 & 1.00 & 0.22 & 0.24 & 0.36 \\
        Typology-5 & -0.04 & 0.04 & 0.37 & 0.22 & 1.00 & 0.20 & 0.42 \\
        Typology-6 & 0.00 & 0.02 & 0.16 & 0.24 & 0.20 & 1.00 & 0.23 \\
        Typology-7 & -0.17 & 0.07 & 0.55 & 0.36 & 0.42 & 0.23 & 1.00 \\
    \bottomrule
  \end{tabular}
  \begin{tablenotes}
        \item[1]\textbf{\hspace{-0.3cm} Typology-1}: I currently have a Facebook account 
        \item[2]\textbf{\hspace{-0.3cm} Typology-2}: I have more than one Facebook account
        \item[3]\textbf{\hspace{-0.3cm} Typology-3}: At some point, I have deactivated my Facebook account 
        \item[4]\textbf{\hspace{-0.3cm} Typology-4}: At some point, I have permanently deleted my Facebook account
        \item[5]\textbf{\hspace{-0.3cm} Typology-5}: At some point, I have voluntarily taken a break from Facebook for a week or more
        \item[6]\textbf{\hspace{-0.3cm} Typology-6}: At some time, I have used software to limit my Facebook usage
        \item[7]\textbf{\hspace{-0.3cm} Typology-7}: At some point, I have deleted the Facebook app from my phone
        \item[8]
    \end{tablenotes}
    \caption{Correlation matrix for non/use questions}
    \label{tab:correlation}
\end{table}

\section{Factor Loadings}

The following tables show all questions and factor loadings used to derive the pyschometric predictors in Section 4.2.

\begin{table}[h]
    \centering
    \Small
    \begin{tabular}{>{\raggedright}p{5cm}ccc}
    \toprule
       Question & Salience, Tolerance \& Mood & Withdrawal \& Conflict & Relapse\\
                & ($\alpha=.83$)              & ($\alpha=.84$)         &\\    
    \midrule
        Spent a lot of time thinking about FB use & 0.83 & &\\
        Felt an urge to use FB more and more & 0.85 & &\\
        FB use to forget about personal problems & 0.57 & & 0.35\\
        Became restless if FB use prohibited & & 0.96 &\\
        FB use had a negative impact on job/studies & & 0.97 & \\
         Tried to cut down FB use without success & & & 0.84\\
    \bottomrule
  \end{tabular}
    \caption{BFAS: Factor analysis.}
    \label{tab:fa:bfas}
\end{table}

\begin{table}[h]
    \centering
    \Small
    \begin{tabular}{>{\raggedright}p{7cm}ccc}
    \toprule
       Question & Connectedness & Daily Routine & Pride\\
         & ($\alpha=.86$)              & ($\alpha=.93$)         &\\  
    \midrule
        I feel out of touch when I haven't logged onto FB for a while & 0.64 & 0.35 & \\
        I feel I am part of the FB community & 0.66 & 0.39 & 0.38 \\
        I would be sorry if FB shut down & 0.71 & 0.35 & 0.30 \\      
        FB is part of my everyday activity & 0.41 & 0.86 &\\   
        FB has become part of my daily routine & 0.58  & 0.65 &\\                 
        I am proud to tell people I am or was on FB & 0.35 &   & 0.89\\          
    \bottomrule
  \end{tabular}
    \caption{FBI: Factor analysis}
    \label{tab:fa:fbi}
\end{table}

\begin{table}[h]
    \centering
    \Small
    \begin{tabular}{>{\raggedright}p{7cm}cc}
    \toprule
       Question & Sense of Negative Agency & Sense of Positive Agency\\
       & ($\alpha=.84$)              & ($\alpha=.79$) \\  
    \midrule
        I am just an instrument in the hands of something else & 0.57 & -0.21\\ %agency-24         
        My actions just happen without my intention & 0.70 &\\ %agency-25         
        Consequences of my actions don't logically follow my actions & 0.71 &\\ %agency-27          
        My movements are automatic - my body simply makes them & 0.64 &\\ %agency-28          
        The outcomes of my actions generally surprise me & 0.75 &\\ %agency29         
        Nothing I do is actually voluntary & 0.50 &\\ %agency-32         
        I feel like I am a remote-controlled robot & 0.68 &\\ %agency-33         
        I am in full control of what I do &      &    0.60\\  %agency-11
        I am the author of my actions &  -0.23    &    0.74\\  %agency-26
        Things I do are subject only to my free will &      &    0.55\\  %agency-30
        The decision whether and when to act is within my hands &      &    0.63\\  %agency-31
        My behavior is planned by me from the beginning to the end &      &    0.55\\  %agency-34
        I am responsible for everything that results from my actions &      &    0.67\\  %agency-35
    \bottomrule
  \end{tabular}
    \caption{Sense of Agency: Factor analysis}
    \label{tab:fa:soa}
\end{table}

\begin{table}[h]
    \centering
    \Small
    \begin{tabular}{>{\raggedright}p{6cm}>{\raggedright}p{1.3cm}>{\raggedright}p{1.3cm}>{\raggedright}p{1.2cm}>{\raggedright\arraybackslash}p{2cm}}
    \toprule
       Question & Personal integration & Social integration & Escape & Affective Gratification\\
       & ($\alpha=.84$)              & ($\alpha=.91$) & ($\alpha=.81$) & ($\alpha=.91$) \\  
    \midrule
        I use FB because it makes me ease off & 0.50 & & 0.31 &\\ %leiner-9                 
        I use FB  to inform myself about certain topics & 0.57 & 0.31 & & 0.31\\ %leiner-11
        I use FB to receive advice and recommendations & 0.61 & & &\\ %leiner-12      
        I use FB to express who I am & 0.72 & & &\\ %leiner-31                           
        I use FB to share my views and opinions & 0.73 & & &\\ %leiner-32                           
        I use FB to keep in touch with friends &     &   0.89 & &\\ %leiner-18
        I use FB to exchange with my friends and family &  0.31   &   0.70 & &\\ %leiner-19
        I use FB because I am bored  &     &       &    0.68 &\\ %leiner-1            
        I use FB to occupy myself &     &       &    0.81 &\\ %leiner-2          
        I use FB because it is fun & 0.39     & 0.35      &       &     0.61\\ %leiner-23  
        I use FB because it is entertaining & 0.34     & 0.33       &       &     0.84\\ %leiner-24   
    \bottomrule
  \end{tabular}
    \caption{Leiner's Scale: Factor analysis}
    \label{tab:fa:leiner}
\end{table}

\end{document}